\documentclass[10pt,a4paper,aps,pra,reprint,showpacs]{revtex4-1}
\pdfoutput=1
\usepackage[utf8x]{inputenc}
\usepackage{ucs}
\usepackage{amsmath}
\usepackage{amsfonts}
\usepackage{amssymb}
\usepackage{makeidx}
\usepackage{booktabs}
\usepackage{lipsum}

\usepackage{pdfpages}
\usepackage{afterpage}

\usepackage{color}
\definecolor{light-gray}{gray}{0.55}

\usepackage{microtype}

\usepackage{graphicx}

\renewcommand{\dag}{^{\dagger}}
\newcommand{\exv}[1]{ \langle #1 \rangle }

\newcommand{\bra}[1]{ \langle #1 \rvert }
\newcommand{\ket}[1]{ \lvert #1 \rangle}

\newcommand{\pfrac}[2]{\frac{\partial #1}{\partial #2}}

\begin{document}

\begin{abstract}
We present a theoretical proposal for a microwave driven circuit composed of superconducting resonators and qubits which shows a bistable behaviour, and we present a simple mechanism that allows single- or few-photon microwave pulses to work as Set- and Reset-signals that switch the circuit between its stable modes. The resulting system constitutes an ultra-low-energy Set-Reset flip-flop, and we show that its memory lifetime far exceeds the lifetime of states stored in any of its separate components.
\end{abstract}

\date{\today}
\author{Christian Kraglund Andersen}
\thanks{E-mail: ctc@phys.au.dk}
\affiliation{Department of Physics and Astronomy, Aarhus University, DK-8000 Aarhus C, Denmark}
\author{Klaus Mølmer}
\affiliation{Department of Physics and Astronomy, Aarhus University, DK-8000 Aarhus C, Denmark}

\title{Circuit QED flip-flop memory with all-microwave switching}


\maketitle

\section{Introduction}

The framework of optical cavity quantum electrodynamics (cavity QED) has recently been implemented in microfabricated electrical system of superconducting resonators and Josephson junctions establishing the field of circuit QED (cQED) \cite{Wallraff:2004oh, PhysRevA.75.032329, dicarlo2009demonstration, you2011atomic}. In particular, in cQED the strong coupling regime has been realized for various implementations of superconducting qubits and microwave resonators \cite{Wallraff:2004oh, PhysRevLett.105.060503, PhysRevB.78.180502, PhysRevLett.107.240501}. While both optical cavity QED and microwave cQED are contestants for succesful implementation of quantum information protocols, there is also growing awareness of the use of the same systems in classical information processing devices. Optical devices in cavity QED have thus recently made a tremendous progress towards ultra-low-power all-optical logical elements \cite{PhysRevA.80.045802, PhysRevLett.105.040502, PhysRevA.84.043821, albert2011cavity, PhysRevLett.108.227402, chen2013all, PhysRevLett.111.203002, PhysRevA.89.023806, santori2014quantum}.
Similarly, implementations for quantum switches \cite{PhysRevB.78.104508,PhysRevA.80.014301} and single-microwave-photon transistors \cite{PhysRevLett.111.063601,manzoni2013single} have been proposed for cQED, mimicking and even surpassing the progress realized with optics.

In the field of microwave photonics, all-microwave logical elements constitute a standing goal \cite{capmany2007microwave,yao2009microwave}. In optical photonics on-chip all-optical switches require ${\sim}10^5$ photons per switch \cite{liu2010ultra}. Ultra-low-energy microwave logical systems working in the few-photon regime of cQED would, thus, greatly outperform current state-of-the-art photonics with respect to minimizing the switching energy. A few-photon photonic device would also be able to temporarily store measurement results from quantum information experiments \cite{PhysRevA.75.032329, dicarlo2009demonstration, PhysRevLett.109.050507} without amplification of microwave signals to levels detectable outside the cryogenic environment in which typical quantum information experiments are performed. This would particularly benefit the application of error correction schemes such as the surface code \cite{PhysRevA.86.032324}.

\begin{figure}[t]
\includegraphics[width=0.99\columnwidth]{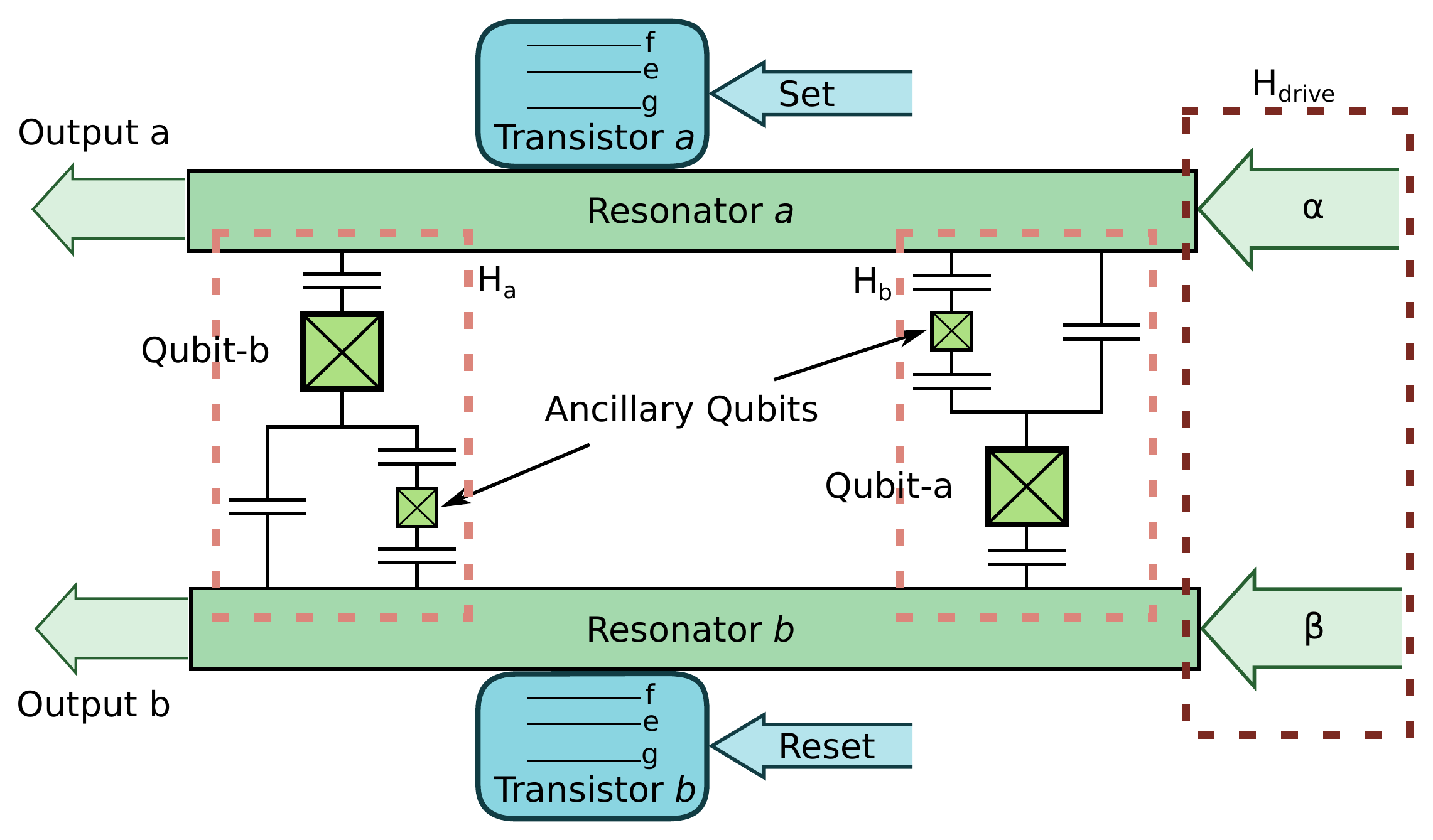}
\caption{We schematically illustrate a system with coupled qubits and microwave resonators implementing the Hamiltonian in Eq. \eqref{eq:fullH}. Qubit-a and qubit-b are resonant with microwave resonators $a$ and $b$, respectively. The states of the two ancialla qubits in the set-up adiabatically follow the states of the resonators and qubits-$a$ an -$b$, and mediate the desired coupling between the systems. The classically driven resonators are coupled to the $e$-$f$ transitions of three-level transmons. When Set- and Reset-signals are applied on the $g$-$e$ transitions, the transmons serve as transistors and control the resonator field. } \label{fig:sketch}
\end{figure}

A classical Set-Reset flip-flop system is the simplest possible memory system and consists of two inputs and two outputs. The output logical states depend on the history of input signals: When a signal pulse arrives at Set the $a$-output is set to 0 and the $b$-output is set to 1, until a signal pulse arrives at Reset, and the $a$-output is set to 1 and the $b$-output is set to 0. The switching between the logical states of a qubit implements a microscopic flip-flop-device and Refs. \cite{chen2013all, manzoni2013single, chang2007single} propose to extend the qubit to a three-level system constituting a photonic transistor for the field modes and thus implementing a few photon input and output device. Superconducting qubits are, however, hampered by fast relaxation and logical memories that rely on qubit-states coupled to microwave-field modes are currently limited to memory times on the order of 10 $\mu$s. This suggests to investigate the use of more complex systems of coupled superconducting components. Elaborate qubit designs, e.g. with fluxonium qubits \cite{pop2014coherent}, or much more complex multi-qubit designs, e.g., the application of surface codes \cite{PhysRevA.86.032324}, may also offer very long quantum and classical storage. In this work, however, we aim for a classical flip-flop memory device that may be controlled by ultra-weak, few-photon microwave pulses, and to this end we shall apply a combination of microwave resonators and qubits, that feature bistable behaviour already at the few photon level, and which shows lifetimes much longer than the ones of the individual components.

\section{Description of the device}

Our goal is to create a flip-flop system, where the two logical memory states are associated with the coherent excitation of one or the other of two weakly driven microwave resonators. A damped cavity subject to continuous driving assumes a coherent steady state excitation, and using the framework and properties of existing components in cQED, we propose in Fig.1 a bistable device that switches between such states when subject to few-photon Set and Reset pulses. The essential mechanism controlling the coherent state amplitudes in the resonators is the frequency shift or splitting of the resonator modes when they are coupled to two- and three-level systems.

In cQED, Josephson junctions form the basis for so-called transmons, where the combined charging energy and tunneling Hamiltonian leads to discrete energy eigenstates that define the qubits and three-level quantum systems \cite{supp, PhysRevA.76.042319}. In Fig. \ref{fig:sketch}, two such three-level transmons, $ta$ and $tb$, are shown in the top and bottom of the figure. They are strongly coupled to microwave resonators with resonance frequencies $\omega_a,\,\omega_b$ via their first and second excited ($e,\,f$) states, while the transition between the ground ($g$) and first excited states is reserved for control by external set- and reset field pulses. The transistor coupling Hamiltonian is given by  ($\hbar = 1$)
\begin{equation}
H_{t} = g_{ta}\, \ket{f}_{ta}\bra{e} \, a + g_{tb}\, \ket{f}_{tb}\bra{e} \, b + \text{H.c.}, \label{eq:transitorham}
\end{equation}
where $a$ and $b$ are the annihilation operators for photons in resonator $a$ and resonator $b$. Both resonators are driven by classical external fields,
\begin{align}
H_{drive} = \alpha (a + a\dag ) + \beta (b+b\dag ), \label{eq:Hdrive}
\end{align}
and we choose $\alpha \, (\beta) = \sqrt{\exv{n_{a\,(b)}} }\kappa_{a\,(b)}/2$, where $\exv{n_{a\,(b)}}$ are the target steady-state photon numbers and $\kappa_{a\,(b)}$ are the photon-loss rates of the resonators.

Consider first the interaction between resonator $a$ and transistor $ta$. If $ta$ is in the state $g$, $H_t$ vanishes and the resonator mode is only subject to $H_{drive}$ and cavity damping which results in a steady state coherent field in the resonator. If a $\pi$-pulse on the $g$-$e$ transition excites $ta$ into state $e$, $H_t$ induces a vacuum Rabi-splitting of the resonator frequency of magnitude $g_{ta}$, and if this is  larger than other coupling strengths and decay rates in the system, it will make $H_{drive}$ in (\ref{eq:Hdrive}) off-resonant. The resonator field will then decay into the vacuum state. Decay of $ta$ back to $g$ will, however, bring the resonator back on resonance with the driving field, and the coherent state will reappear: although it is stored in the resonator field, the memory of the $\pi$-pulse excitation is governed by the transistor excited state lifetime.

The $\pi$-pulses need, in this scheme, to be injected by separate lines connected to $ta$ and $tb$. There are no requirements on how the pulses are created as long as the pulse-areas correspond to a fast $\pi$-rotation on the $g$-$e$ transition. We note, however, that symmetrically shaped single-photon wave packets \cite{pechal2013microwave, PhysRevB.84.014510, srinivasan2013time, PhysRevLett.110.107001} would meet the desire of having as few photon as possible.

If one could maintain the state of resonator $a$ after the transistors decay into $g$, we would already have a flip-flop, however, this is not the case with only a transistor and a resonator. Thus, to extend the time over which the cavity field memory states recall the excitation of both $ta$ and $tb$, we propose the elaborate system design in Fig. \ref{fig:sketch} where the resonators are coupled to each other via two qubits. The two ancillary qubits shown are adiabatically eliminated but they are crucial to mediate the desired effective interaction Hamiltonian \cite{supp},
\begin{equation}
\begin{aligned}
H_a = \chi_a\, a\dag a \, (b\dag \sigma_{b,-} + b\sigma_{b,+}), \\ H_b = \chi_b\, b\dag b \, (a\dag \sigma_{a,-} + a\sigma_{a,+}),
\end{aligned} \label{eq:H}
\end{equation}
where $\sigma_{a\,(b),-}$ and $\sigma_{a\,(b),+}$ are the lowering and raising operators for qubit-a and qubit-b in the figure.

With the Hamiltonian $H_a$ we achieve that a non-vanishing field in resonator $a$ gives rise to a strong Jaynes-Cummings coupling between resonator $b$ and qubit-$b$. We already discussed how such a coupling causes a splitting of the resonator frequency, so as a consequence of the excitation in resonator $a$ the driving of resonator $b$ now becomes off-resonant and inefficient. The set-up is symmetric, and a field in resonator $b$ similarly controls the strength of the coupling between resonator $a$ and its qubit and hereby prevents its resonant driving. The switching pulses still work as described above: e.g., a switching $\pi$-pulse exciting transistor $ta$ stops the driving of resonator $a$. The field in resonator $a$ decays on a time scale of $1/\kappa_a$, allowing the field to build up in resonator $b$, and now, even after $ta$ has decayed (assuming a lifetime $T_{ta,1}> 1/\kappa_a$), resonator $a$ cannot be excited because of $H_b$ and the coherent state in resonator $b$. The coherent state in resonator $b$ is maintained until a Reset pulse on $tb$ shifts this resonator off resonance, the field decays, and now the coherent state builds up in resonator $a$ and prevents excitation of resonator $b$ also after $tb$ has returned to its ground state. 

Summing up, the memory states are a driven coherent state in one resonator and the vacuum state in the other and vice versa. By a non-linear cross-coupling between the resonators mediated by qubits, the field in one resonator prevents the excitation of the other one and preserves the memory states for times longer than both the resonator and qubit lifetimes. The Set and Reset pulses applied to the transistor $ta$  ($tb$) works by putting resonator $a$ ($b$) off resonance. This causes the field to decay and the other field to grow and, thus, the system switches between the memory states. The scheme is robust against decay and decoherence, since the switching and the preservation of the memory states relies on the interaction induced vacuum-Rabi splitting, and while dephasing or noise on the qubits may change the coupling strengths and introduce damping effects in (\ref{eq:H}), the precise interaction is not crucial, as their role is not to mediate but to prevent excitation of the resonator fields.

To elaborate briefly on how $H_a$ and $H_b$ in \eqref{eq:H} are implemented we consider the circuit illustrated in Fig. \ref{fig:sketch} (see further details in Ref. \cite{supp}). We apply first- and second-order perturbative expressions for the high frequency capacitative energies to include both the direct first-order couplings between the qubits and the resonators as well as the second-order coupling through the ancillary qubits. We then adiabatically eliminate the ancillary qubits and the resulting Hamiltonians become \cite{supp}:
\begin{equation}
\begin{aligned}
H_a' =&\, \Big(g_a + \big(\chi_a^{(1)} - \chi_a^{(2)}a\dag a\big)\; a\dag a\Big) (b\dag \sigma_{b,-} + b\sigma_{b,+}), \\ H_b' =&\, \Big(g_b + \big(\chi_b^{(1)} - \chi_b^{(2)}b\dag b\big)\; b\dag b\Big) (a\dag \sigma_{a,-} + a\sigma_{a,+}).
\end{aligned} \label{eq:H1}
\end{equation}
The coupling parameters depend on $\Delta = \omega_a - \omega_b$, and tuning the frequency of the qubits \cite{PhysRevA.76.042319,PhysRevLett.106.030502} as well as the resonators \cite{palacios2008Tunable,sandberg2008tuning} we can obtain $g_a = g_b = 0$, and effectively obtain Eqs. \eqref{eq:H} with $\chi_a=(\chi_a^{(1)} - \chi_a^{(2)}a\dag a)$ and $\chi_b=(\chi_b^{(1)} - \chi_b^{(2)}b\dag b)$. From the proposed implementation, we also get a cross-Kerr coupling between the resonators \cite{PhysRevA.84.012329}. After cancellation of the lower order terms $\propto g_a, g_b$, the fourth order terms, $\chi_{a \, (b)}^{(1)}$, are dominant. Higher order corrections to the coupling, $\chi_{a \, (b)}^{(2)}$, arise from saturation of the adiabatically eliminated ancillary qubits, and we have included them in our analysis to show that even though they reduce the effective values of $\chi_{a \, (b)}$, they do not affect the bistable behaviour in a detrimental way.

We implement the Hamiltonians in \eqref{eq:H1} and include all contributions, such that the full Hamiltonian is given by
\begin{align}
H =& \phantom{+}\, \big(\chi_a^{(1)} - \chi_a^{(2)}a\dag a\big)\; a\dag a\; (b\dag \sigma_{b,-} + b\sigma_{b,+}) \nonumber \\
& {+}\, \big(\chi_b^{(1)} - \chi_b^{(2)}b\dag b\big)\; b\dag b\; (a\dag \sigma_{a,-} + a\sigma_{a,+}) \nonumber \\
& {+}\, \chi^{(ab)}\, a\dag a\; b\dag b \quad + H_t + H_{drive}. \label{eq:fullH}
\end{align}
Using realistic numbers for the implementation \cite{supp} the cross Kerr term yields $\chi^{(ab)} = 2\pi \times 0.07$ MHz, and we can achieve an effective coupling strength between resonator $b$ and its qubit of around $\tilde{g}_b = \chi_a \exv{a\dag a}  = 2\pi \times 7.1$ MHz, for $ \exv{a\dag a} = 8,\ \exv{b\dag b} = 0$, while for resonator $a$ we can achieve $\tilde{g}_a =\chi_b \exv{b\dag b} = 2\pi \times 7.6$ MHz with $\exv{b\dag b} = 8,\ \exv{a\dag a} = 0$. In order to have a strong coupling, $\tilde{g}_i^2/\kappa\gamma \gg 1$, we use a lifetime of the qubits $T_{qa\,(qb),1} = 1/\gamma = 12 \, \mu$s and a decay-rate of the cavities of $\kappa = 2\pi \times 0.1$ MHz. With these parameters, the power needed to drive the resonators is a mere $2\times 10^{-17}$ W. Also, with these resonator-parameters we must require a transistor qubit lifetime, $T_{ta, (tb),1}$ of around 20 $\mu$s and to achieve this, state of the art transmon \cite{PhysRevA.76.042319,PhysRevLett.106.030502} or Xmon \cite{PhysRevLett.111.080502} qubits must be used.

\section{Quantum trajectory simulations}

The average time-evolution of the system is governed by the master equation \cite{carmichael2002statistical}
\begin{align}
\pfrac{\rho}{t} =&\, i[\rho,H] + \sum_{k\in \{a,b\}} \Big(\frac{\kappa_k}{2} \, (2\,k\rho k\dag -k\dag k \rho -\rho k\dag k) \nonumber \\
&+ \frac{\gamma_{t}}{2} \, (2\, \sigma_{tk,-} \rho \sigma_{tk,+} - \sigma_{tk,+} \sigma_{tk,-} \rho - \rho \sigma_{tk,+} \sigma_{tk,-}) \nonumber \\
&+ \frac{\gamma}{2} \, (2\, \sigma_{k,-} \rho \sigma_{k,+} - \sigma_{k,+} \sigma_{k,-} \rho - \rho \sigma_{k,+} \sigma_{k,-}) \Big) \label{eq:master}
\end{align}
with $\kappa_a$, $\kappa_b$, $\gamma_t$ and $\gamma$ being the decay rate of the cavities, the exited states of the transistor transmons and the qubits. We will assume $\kappa_a = \kappa_b = \kappa$. Instead of directly solving Eq. \eqref{eq:master}, which will yield an average over the bistable behaviour of the device, we apply Monte Carlo wave function (MCWF) simulations. Such simulations reproduce on average the result of the master equation. The simulations apply propagation by a non-unitary Schrödinger equation \cite{carmichael2002statistical, wiseman2010quantum, PhysRevLett.68.580}, $\frac{d}{dt} \ket{\psi(t)} = -\Big( \sum_{\mu} c_{\mu}\dag c_{\mu}/2 + iH \Big) \ket{\psi(t)} $, interrupted by the application of quantum jumps $\ket{\psi} \rightarrow c_{\mu}\ket{\psi}$, where $c_{\mu}$ are the jump operators for all decay channels, e.g., $\sqrt{\kappa}\, a$ for resonator $a$ and $\sqrt{\gamma}\, \sigma_{a,-}$ for the qubit-a etc. We refer to a renormalized solution $\ket{\psi(t)}$ as a quantum trajectory and the observable mean values calculated from a single quantum trajectory is what one would infer as system variables if one had access to a readout of all decay channels. More generally, systems subject to partial or inefficient monitoring are described by density matrices obeying stochastic master equations \cite{wiseman2010quantum}. Photon leakage from the resonators is the dominant and most frequent jump process and detection of just a few of these photons is enough to distinguish the two states of the system. We thus expect that the corresponding stochastic master equation is well represented by the MCWF pure state dynamics.

\begin{figure}[b]
\includegraphics[width=0.99\columnwidth]{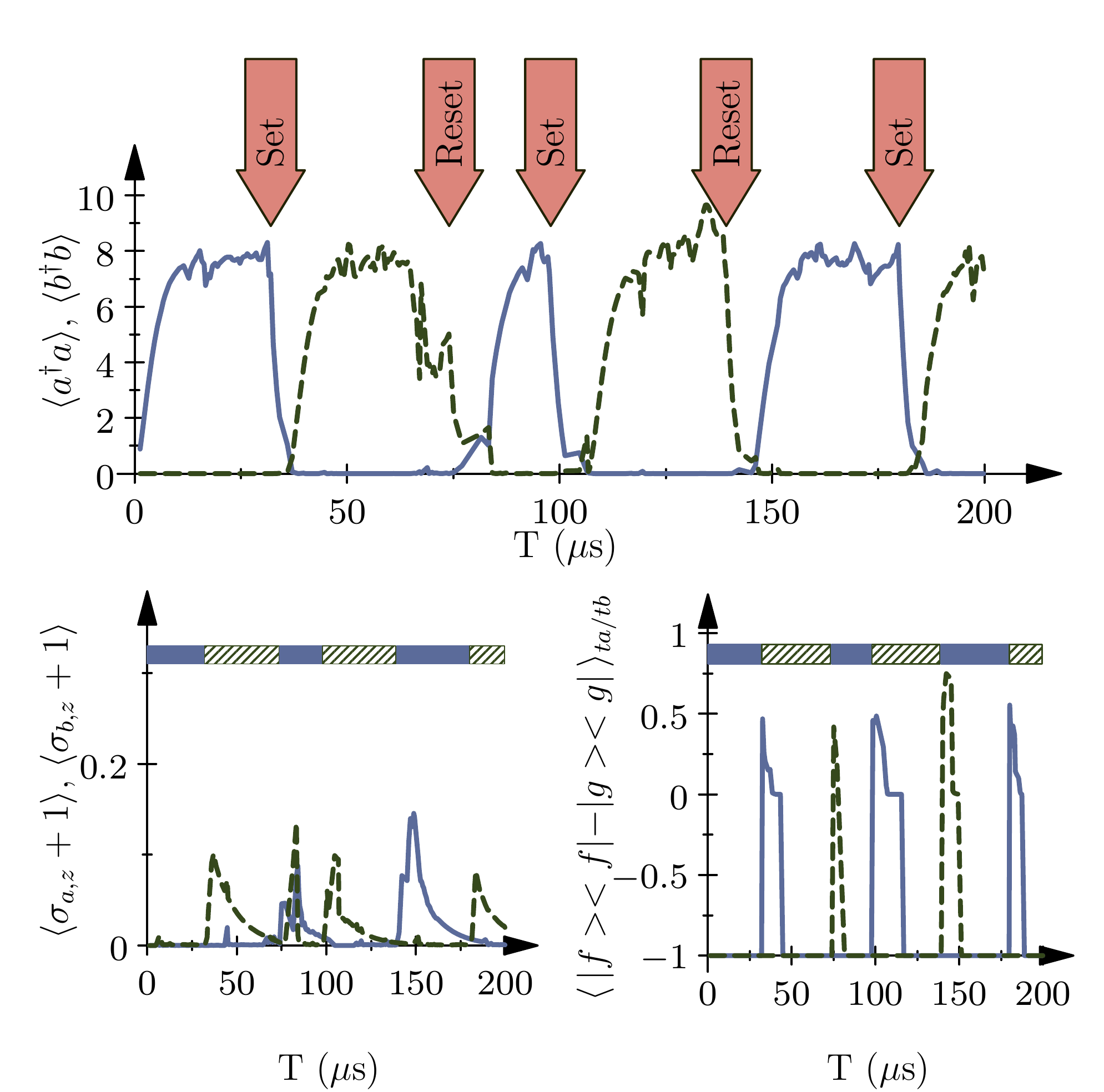}
\caption{Upper panel: A single quantum trajectory with Set and Reset pulses applied at the times indicated. The solid (blue) curve is $\exv{a\dag a}$ and the dashed (green) curve is $\exv{b\dag b}$. Lower left panel: Population of qubit-a (solid blue curve) and qubit-b (dashed green curve). Lower right panel: Expectation value of $\ket{f}_{ta}\bra{f} - \ket{g}_{ta}\bra{g}$ (solid blue curve) and $\ket{f}_{tb}\bra{f} - \ket{g}_{tb}\bra{g}$ (dashed green curve). The bar in the top of both lower panels indicates that resonator $a$ (blue) or resonator $b$ (green hatched) is excited. The parameters used in the simulations are $(\chi_a^{(1)},\, \chi_a^{(2)},\, \chi_b^{(1)}, \chi_b^{(2)},\, \chi^{(ab)} ) = 2\pi \times (0.98,\,0.011,\,1.04,\,0.012,\,0.07)$ MHz and the resonator frequencies are $\omega_a = 2\pi \times 7$ GHz and $\omega_b = 2\pi \times 5$ GHz. We use a decay-rate of the cavities at $\kappa = 2\pi \times 0.1$ MHz and a lifetime of qubit-a and -b of $12 \, \mu$s. For both transitors we assume $g_{ta\, (tb)} = 2\pi \times 30$ MHz and lifetimes of 20 $\mu$s.}  \label{fig:flipflop}
\end{figure}

In Fig. \ref{fig:flipflop} we present a quantum trajectory simulation of the device. The upper panel of Fig. \ref{fig:flipflop} shows that the flip-flop operates as we expect. At first we start driving resonator $a$ and shortly after we start driving resonator $b$. No population, however, appears in resonator $b$ due to the induced split of the resonance frequency. After a short time ($\sim 32 \, \mu$s) we apply a $\pi$-pulse on the $g$-$e$-transition of transistor qubit $a$ -- this is our Set signal and the population in resonator $a$ will decay as the drive field is no longer resonant due to the coupling to the $e$-$f$-transition of the transistor qubit. Meanwhile resonator $b$ is resonantly coupled to the drive, and when the transistor decays to the ground-state at $\sim 50 \, \mu$s, resonator $b$ has become excited and induces a strong coupling in $a$ such that it remains empty. The same procedure is repeated with further Set- and Reset-pulses in the figure.

In the lower left panel of Fig. \ref{fig:flipflop} we see that while resonator $a$ is populated and $H_a$ is 'active', qubit-b remains in the ground state, while when resonator $b$ is populated qubit-b undergoes excitation dynamics during the switching process. This excitation in qubit-b is not important since the Hamiltonian, $H_b$, responsible for keeping resonator $a$ empty and resonator $b$ excited does not contain qubit-b operators. The lower right panel of Fig. \ref{fig:flipflop} shows the corresponding transient evolution of the transistor population of states $\ket{g}$ and $\ket{f}$, triggering the filling and emptying of the resonators.

While the master equation \eqref{eq:master} cannot be solved analytically, an approximate analysis in \cite{supp} confirms that the steady state density matrix is a mixture of two solutions with a non-vanishing field in one or the other resonator, accompanied by a weak excitation of one or the other qubit. The steady state analysis of the correlations between the qubit and field states thus confirms the bistable solutions of the MCWF simulations.

\begin{figure}[b]
\includegraphics[width=0.99\columnwidth]{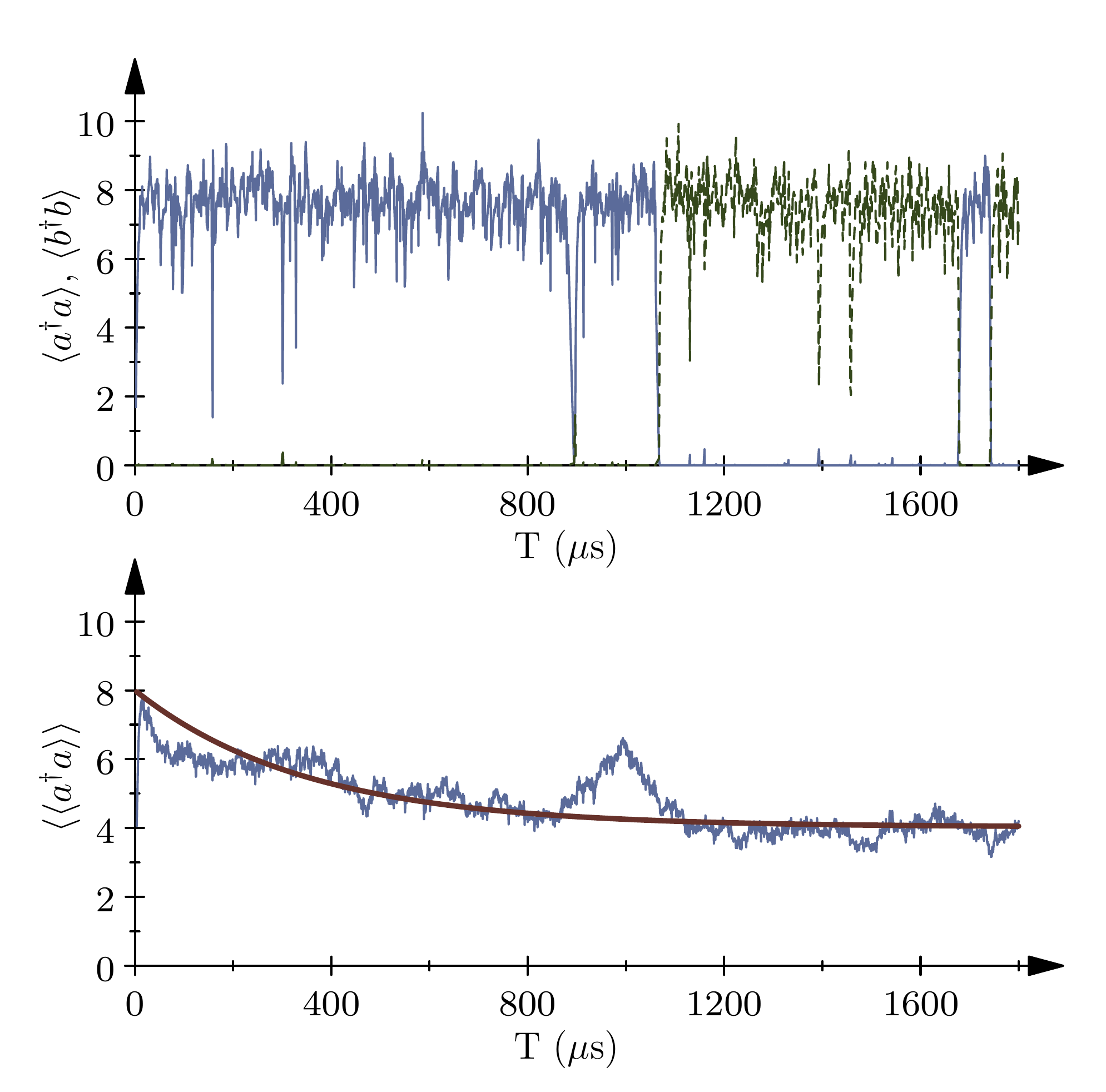}

\caption{Upper figure: A single trajectory with no switching pulses applied. The solid (blue) curve is $\exv{a\dag a}$ and the dashed (green) curve is $\exv{b\dag b}$. Lower figure: An ensemble averaged mean of $\exv{a\dag a}$ over 30 trajectories (thin light blue) and an exponential fit (thick dark red) with the decay-time 347 $\mu$s. The parameters used are the same as in Fig. \ref{fig:flipflop}.}  \label{fig:no_flip}
\end{figure}

The trajectory in the upper panel of Fig. \ref{fig:no_flip} shows that in the absence of Set- and Reset-pulses, the flip-flop undergoes spontaneous state changes, and we estimate the rate of such erroneous switches to be about one every 600 $\mu$s. We have further quantified the behaviour over many realizations in the lower panel of Fig. \ref{fig:no_flip}, where  we have used $N=30$ trajectories to generate the ensemble averaged mean photon number in the $a$-resonator when no Set- and Reset-pulses are applied. Fitting the relaxation of this mean value, we find a memory time of $347.7 \, \mu$s with an uncertainty around $50\, \mu$s, which is over 2 orders of magnitude longer than the bare cavity lifetime at 1.5 $\mu$s and also much longer than the qubit lifetime of 12 $\mu$s. Note at this point, that if an error occurs, already after a subsequent Set- and Reset-pulse the device returns to the desired memory state.

\section{Estimates of performance}

\begin{figure}[t]
\includegraphics[width=0.99\columnwidth]{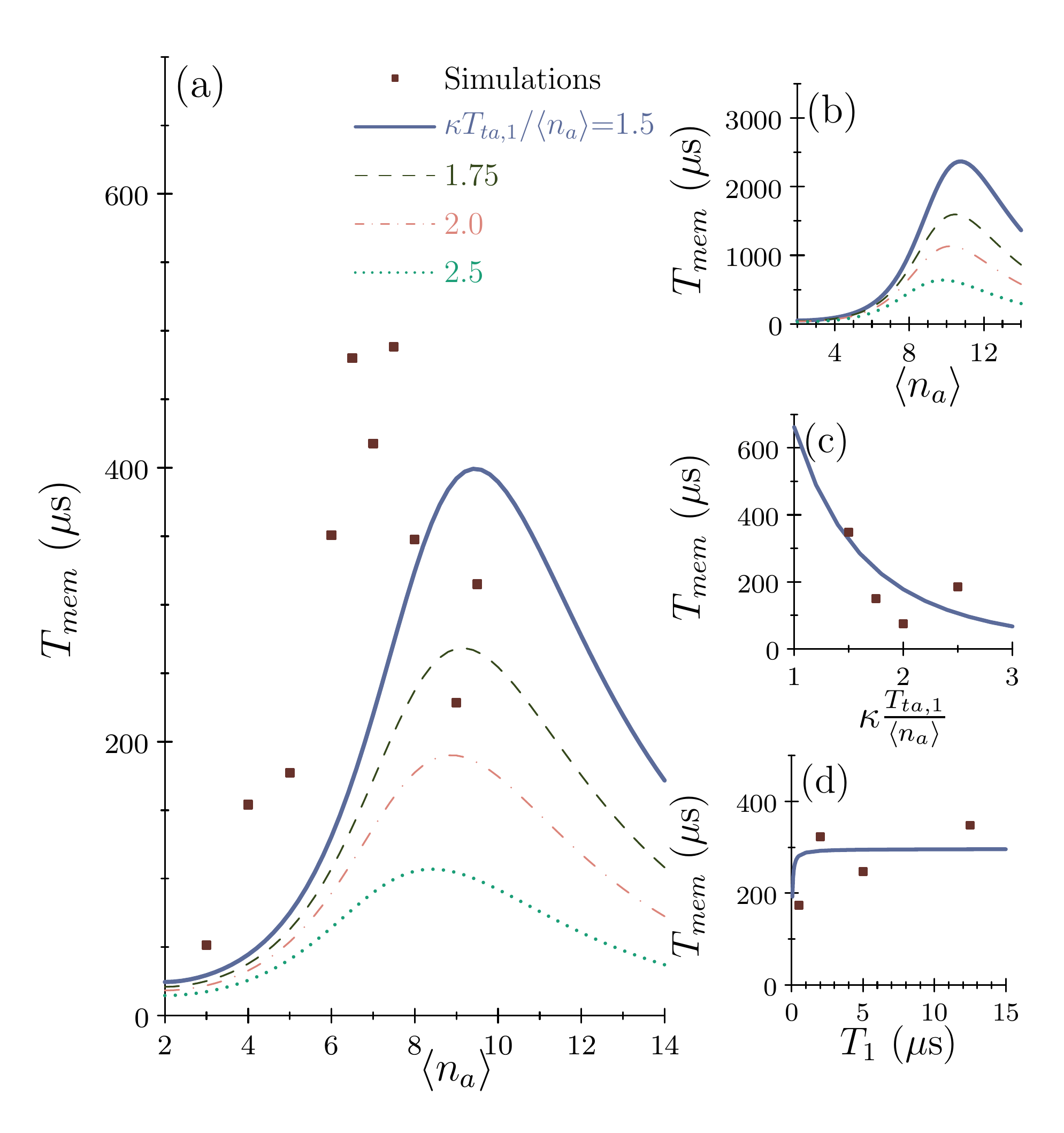}

\caption{Memory time simulated using parameters specified in Fig. \ref{fig:flipflop} and calculated using Eq. \eqref{eq:t-estimate} and in (d) using the qubit excitation estimate as well. All simulated results contain an uncertainty on the order of 10 $\mu$s. In (a) we vary the target photon number in the resonators while keeping the fraction $\kappa T_{ta,1}/ \exv{n_a}$ constant at the values specified. In the simulations represented by the square symbols, we use $\kappa T_{ta,1}/ \exv{n_a} = 1.5$. In (b) we show the analytical estimates as in (a) with the same parameters except $T_{ta,1} = 40 \, \mu$s. In (c) we vary $\kappa$ while keeping $\exv{n_a} = 8$ and $T_{ta,1} = 20 \, \mu$s and in (d) we study the memory time for different values of the qubit-$a$ and -$b$ life times.}  \label{fig:estimates}
\end{figure}

To supplement our numerical simulations of the functioning of the device, we shall derive approximate expressions that can reveal dependencies on the component parameters and indicate the prospects for its optimization and improvement. We estimate the memory time of the flip flop in the resonator $a$ state as
\begin{align}
\frac{1}{{T}_{mem}} = \sum_{n} e^{-\exv{n_a}}\frac{\exv{n_a}^n}{n!} \, \frac{(2\beta)^2}{\kappa^2 + (\chi_a^{(1)}n - \chi_a^{(2)}n^2)^2}  \, \kappa   \label{eq:t-estimate}
\end{align}
which is the feeding rate of photons in the off-resonant resonator $b$ weighted over Poisson distributed number states occupying resonator $a$. This yields ${T}_{mem}=340\, \mu$s for the parameters used, which is in qualitative agreement with our simulations. Using Eq. \eqref{eq:t-estimate} we have calculated the memory time for different values of the mean excitation of resonator $a$. When varying $\exv{n_a}$, we change $\kappa$ in order to keep the ratio $\kappa / \exv{n_a}$ constant. Estimates from \eqref{eq:t-estimate} for different values of $\kappa T_{ta,1} / \exv{n_a}$ are shown as curves in Fig. \ref{fig:estimates} (a). The numerically simulated memory times for $\kappa T_{ta,1} / \exv{n_a} = 1.5$ are calculated from exponential fits similar to that of Fig. \ref{fig:no_flip}. We see that our rough estimate of Eq. \eqref{eq:t-estimate} qualitatively reproduce the simulated behaviour. The maximum memory time is fairly well predicted, but the optimal photon number is slightly shifted compared to the simulations. This may be due to an underestimation of the effect of $\kappa$. To achieve a better Set-Reset-performance, $\kappa T_{ta,1} / \exv{n_a}$ must be increased, but we see that this decreases the memory time. In \ref{fig:estimates} (b), however, we see that with a better transistor qubit we can improve the memory time significantly, even with high values of $\kappa T_{ta,1} / \exv{n_a}$. In Fig. \ref{fig:estimates} (c) we compare simulations  with the estimate of \eqref{eq:t-estimate} as a function of $\kappa T_{ta,1} / \exv{n_a}$ by varying $\kappa$.

In Fig. \ref{fig:estimates} (d) we show how the use of worse qubits influence the memory time. For a quantitative analysis of the qubit contribution, we approximate the master equation, neglecting contributions from $H_b$, by coupled equations for the steady-state qubit population in the excited and ground state, $P_{\uparrow / \downarrow}$, and the accompanying field amplitude of each resonator, $\alpha_{\uparrow / \downarrow}$ and $\beta_{\uparrow / \downarrow}$ \cite{Kilin:91},
\begin{align}
\frac{P_\uparrow}{P_\downarrow} = \frac{\frac{4\chi_a^2}{\gamma^2}(|\alpha_{\downarrow}|^4 +|\alpha_{\downarrow}|^2)|\beta_\downarrow|^2}{1 + \frac{4\chi_a^2}{\gamma^2}(|\alpha_{\uparrow}|^4+|\alpha_{\uparrow}|^2) (|\beta_\uparrow|^2 + 1)}
\end{align}
(See detailed derivation in Ref. \cite{supp}). We estimate the qubit contribution to the memory time by adding $P_{\uparrow} \gamma$ to $1/T_{mem}$. Our simulations confirm the qubit effect and the sharp decrease in the memory time when $T_1$ becomes less than $\sim 1$ $\mu$s. Since we are approaching the edge of the strong coupling regime we cannot expect the photon blockade to function well here. To summarize, the dominant contribution to the erroneous flipping between the states originates from the finite probability of having zero photons according to the Poisson distributed population of the resonators. If qubit-a and qubit-b have very short lifetimes, their relaxation will also contribute significantly to the erroneous state changes.

\section{Conclusion and Outlook}

In conclusion we have proposed a scheme for implementing a flip-flop system operating in the few microwave photon regime of cQED. The development is inspired by optical cavity QED, but due to the absence of the long-lived and phase stable states offered by atoms, we use a two-resonator Hamiltonian, where the excitation of one resonator blocks the excitation of the other one. Using realistic parameters we show that with single photon pulses we can switch between two stable states and that the systems memory time far exceeds that of its intrinsic components. This type of memory system is a significant step towards classical microwave logic in cQED. Promising devices have also been proposed and demonstrated with coherent fields, weak non-linearities and feedback \cite{PhysRevLett.109.153602,PhysRevX.3.021013}. In contrast, our proposed device exploits strong vacuum Rabi-splitting and operates in the regime of few photons. The primary limitations of our proposal are set by the life-time of the transistor qubits, but one can expect future superconducting qubits with much longer life-times \cite{PhysRevLett.111.080502,PhysRevB.86.100506,pop2014coherent} to improve the performance.

As a possible extension, avoiding the difficulties of creating $\pi$-pulses, one might imagine that resonator $a$ ($b$) is not resonant with the $e$-$f$ transition of $ta$ ($tb$). Then a cw-signal with driving amplitude $\Omega_d$ detuned from the $g$-$e$ transition by $\Delta_{SR} = \omega_d - \omega_{ge} = \omega_{ef} - \omega_a$ will generate a second order coupling of $\tilde{g}_{ta} = g_{ta} \Omega_d / \Delta_{SR}$. If $\tilde{g}_{ta}$ is much larger than any other coupling strength or decay rate resonator $a$ will decay to the ground state during the duration of the driving signal and we retrieve the same functionality as before. This type of pulses, which typically will consist of ${\sim} 100$ photons, might originate from an elongated dispersive readout of a superconducting qubit \cite{mallet2009single}.

The system that we propose involves more components than a typical cQED memory device, but the added complexity serves a purpose and improves the system characteristics beyond what is currently achievable with single components. We believe that it is worthwhile for a wide range of possible applications to see the complexity not as a disadvantage but, rather, as an opportunity to pursue circuit strategies along with the efforts to improve the performance of individual components.

\section{Acknowledgement}
We thank G. Oelsner and J. Kerchoff for useful inputs and P. Haikka, D. Petrosyan, M.C. Tichy and D. Dasari for feedback on the manuscript. We furthermore acknowledge support from the Villum Foundation and the EU 7th Framework Programme collaborative project iQIT.

\bibliography{bt}

\newpage\null\thispagestyle{empty}\newpage

\onecolumngrid


\section*{Supplementary material}

# Flip-flop memory in circuit QED with all-microwave switching [1]: Supplementary Material

Christian Kraglund Andersen* and Klaus Mølmer
*Department of Physics and Astronomy, Aarhus University, DK-8000 Aarhus C, Denmark*
(Dated: September 22, 2014)

In this Supplementary Material we derive the effective Hamiltonian, implemented by our proposed circuit. We determine the coupling strengths and we identify useful parameter regimes. An approximate factorized ansatz for the qubits is applied to identify average properties and dynamical properties of the device are illustrated by Monte Carlo trajectories.

## I. DERIVATION OF HAMILTONIAN

In Eq. (3) in the article [1], $H_a$ and $H_b$ describe a coupled system of two resonators and two transmon qubits. We consider here the construction of

$$H_a = \chi\, a^\dagger a\, (b^\dagger \sigma_{b,-} + b\sigma_{b,+}). \tag{1}$$

by the circuit illustrated in Fig.1, where $a$ and $b$ are annihilation operators for two driven resonators and $\sigma_{b,\pm}$ are the ladder operators for Qubit-$b$. As we shall show in the following, incorporation of a capacitively coupled ancillary qubit component, Qubit-$anc$, leads to the highly non-linear coupling terms in Eq. (1) ($H_b$ In Eq. (3) in the article [1] is obtained in a similar manner).

We start by writing the Lagrangian of the system. When writing the Lagrangian we have chosen the junctions as the spanning tree and for the node variables at the superconducting islands [2]. At first, we neglect contributions from the resonators – they will be reintroduced later as high frequency gate charge contributions. Summing up we have the Lagrangian

$$\mathcal{L} = \frac{2C_{J,b}\dot\phi_b^2}{2} + \frac{2C_{J,anc}\dot\phi_{anc}^2}{2} + \frac{C_{12}(\dot\phi_b - \dot\phi_{anc})^2}{2} + \frac{C_{a1}\dot\phi_b^2}{2} + \frac{C_{b1}\dot\phi_b^2}{2} + \frac{C_{b2}\dot\phi_{anc}^2}{2} + E_{J,b}\cos\phi_b + E_{J,anc}\cos\phi_{anc}, \tag{2}$$

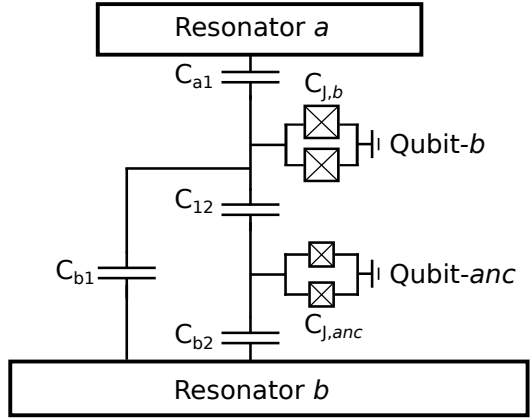

FIG. 1. Schematic figure of two resonators coupled via two transmon qubits. Qubit-$b$ is coupled to resonator $a$, resonator $b$ and Qubit-$anc$, while Qubit-$anc$ is only coupled to resonator $b$ and Qubit-$b$. After ellimination of Qubit-$anc$, we obtain $H_a$ of the main article. Unlike the simplified Fig. 1 in the main article, we have here displayed all circuit elements required to generate the qubits.

---

* E-mail: ctc@phys.au.dk

where the shunted capacitors parallel to the junctions are absorbed into $2C_{j,b}$. We can now find the conjugate variables,

$$q_b = \frac{\partial \mathcal{L}}{\partial \dot{\phi}_b} \qquad q_{anc} = \frac{\partial \mathcal{L}}{\partial \dot{\phi}_{anc}}, \qquad (3)$$

and transform into the two-transmon Hamiltonian

$$H = 4E_{C,b} n_b^2 - E_{J,b} \cos\phi_b + 4E_{C,anc} n_{anc}^2 - E_{J,anc} \cos\phi_{anc} + 4E_I n_b n_{anc} \qquad (4)$$

with $q_k = -2e n_k$ and

$$E_{C,b} = e^2 \frac{C_{12} + C_{b2} + 2C_{J,anc}}{2C_\Sigma^2} \qquad (5)$$

$$E_{C,anc} = e^2 \frac{C_{12} + C_{a1} + C_{b1} + 2C_{J,b}}{2C_\Sigma^2} \qquad (6)$$

$$E_I = e^2 \frac{C_{12}}{C_\Sigma^2}, \qquad (7)$$

where $C_\Sigma^2 = C_{12}(C_{a1} + C_{b1} + 2C_{J,b} + C_{b2} + 2C_{J,anc}) + (C_{b2} + 2C_{J,anc})(C_{a1} + C_{b1} + 2C_{J,b})$.

We now assume coupling to the resonators at anti-nodes of the voltage mode functions, and by including contributions to the gate charge from second-order voltage fluctuations [3] we can replace the charge number, $n_k$ by $n_k - n_{g,k}$, where the gate-charges are given by

$$n_{g,b} = \frac{C_{b1} V_{b,0}}{2e} i(b - b^\dagger) \pm \frac{C_{a1} V_{a,0}}{2e} i(a - a^\dagger) \qquad (8)$$

$$n_{g,anc} = \frac{C_{b2} V_{b,0}}{2e} i(b - b^\dagger), \qquad (9)$$

with the root mean square of the voltage fluctuations in each resonator given by $V_{j,0} = \sqrt{\omega_j/(2C_j)}$, with $C_j$ being the capacitance of resonator $j$. In the limit of $E_J \gg E_C$, the Hamiltonian $H_k = 4E_{C,k} n_k^2 - E_{J,k} \cos\phi_k$ can be approximated $H_k = \sqrt{8E_C E_J} a_k^\dagger a_k - E_C(a_k + a_k^\dagger)^4/12$, with $a_k$ being the annilation operator for a harmonic oscillator. With these operators we can express the charge as $n_k = -i(E_J/8E_C)^{1/4}(a_k - a_k^\dagger)/\sqrt{2}$. If the coupling strengths are much weaker than the anharmonicity of the transmon given by approximately $-E_C$, we can restrict ourselves to the two lowest energy levels and replace $a_k$ by a Pauli lowering operator $\sigma_{k,-}$. The $\pm$-sign in Eq. (8) is obtained if we assume that $C_{b1}$ and $C_{b2}$ connect to the resonator $b$ at integer (half-integer) wavelength separation with the same (opposite) coupling amplitude. This is possible with meandering resonators and is particular manageable using the arms of the Xmon design of the transmon [4].

This now allows us to recast the Hamiltonian in the qubit basis of each transmon and using the rotating wave approximation we end up with the Hamiltonian

$$\begin{aligned} H = & \frac{\Omega_b}{2}\sigma_{z,b} + \frac{\Omega_{anc}}{2}\sigma_{z,anc} + \omega_a a^\dagger a + \omega_b b^\dagger b \\ & + g_{ab}(a^\dagger b + b^\dagger a) + g_{12}(\sigma_{b,+}\sigma_{anc,-} + \sigma_{anc,+}\sigma_{b,-}) \\ & + g_{a1}(a^\dagger \sigma_{b,-} + \sigma_{b,+} a) + g_{b1}(b^\dagger \sigma_{b,-} + \sigma_{b,+} b) \\ & + g_{a2}(a^\dagger \sigma_{anc,-} + \sigma_{anc,+} a) + g_{b2}(b^\dagger \sigma_{anc,-} + \sigma_{anc,+} b), \end{aligned} \qquad (10)$$

where $\Omega_k \approx \sqrt{8E_{C,k} E_{J,k}}$ and the coupling strength are given by

$$g_{ab} = \pm(8E_{C,b} + 4E_I)\frac{C_{b1} C_{a1}}{4e^2} V_{0,a} V_{0,b} \qquad (11)$$

$$g_{12} = 4E_I \beta_{anc} \beta_b \qquad (12)$$

$$g_{a1} = -8E_{c,b} \beta_b \frac{C_{a1} V_{0,a}}{2e} \qquad (13)$$

$$g_{a2} = -4E_I \beta_{anc} \frac{C_{a1} V_{0,a}}{2e} \qquad (14)$$

$$g_{b1} = -4E_I \beta_b \frac{C_{b2} V_{0,b}}{2e} \mp 8E_{c,b} \beta_b \frac{C_{b1} V_{0,b}}{2e} \qquad (15)$$

$$g_{b2} = -4E_I \beta_{anc} \frac{C_{b1} V_{0,b}}{2e} - 8E_{c,anc} \beta_{anc} \frac{C_{b2} V_{0,b}}{2e}. \qquad (16)$$

TABLE I. Component parameters and the resulting effective coupling strengths. We provide two different sets of parameters corresponding to the implementation of $H_a$ and $H_b$ (with $a \leftrightarrow b$) in the main article, respectively. For both sets of parameters we assume a mean photon number of 8 in the resonantly driven resonator. Note that $\Delta = \pm 2\pi \times 1.5$ GHz, has opposite signs for the two parameters sets - this leads to a reduction of the cross-Kerr coupling between the two resonators. $T_1$ for the adiabatically eliminated qubit-$anc$ is set to 50 ns.

|  | $\omega_a = 2\pi \times 7$ GHz<br>$\omega_b = 2\pi \times 5$ GHz | $\omega_a = 2\pi \times 5$ GHz<br>$\omega_b = 2\pi \times 7$ GHz |
|---|---|---|
| $C_{J,1}$ | 45 fF | 32 fF |
| $C_{J,2}$ | 11 fF | 15 fF |
| $C_{a1}$ | 0.90 fF | 3.31 fF |
| $C_{b1}$ | 0.86 fF | 1.70 fF |
| $C_{b2}$ | 9.18 fF | 8.84 fF |
| $C_{12}$ | 3.70 fF | 3.48 fF |
| $\delta$ | $2\pi \times 31.4$ MHz | $2\pi \times 27.5$ MHz |
| $E_{J,b}/E_{C,b}$ | 83 | 93 |
| $E_{J,anc}/E_{C,anc}$ | 24 | 18 |
| $E_{c,b}$ | $2\pi \times 202$ MHz | $2\pi \times 266$ MHz |
| $E_{c,anc}$ | $2\pi \times 545$ MHz | $2\pi \times 456$ MHz |
| $g_{ab}$ | $2\pi \times 0.05$ MHz | $2\pi \times -0.464$ MHz |
| $g_{12}$ | $2\pi \times 199.3$ MHz | $2\pi \times 197.6$ MHz |
| $g_{a1}$ | $2\pi \times -13.1$ MHz | $2\pi \times -46.6$ MHz |
| $g_{a2}$ | $2\pi \times -1.0$ MHz | $2\pi \times -2.53$ MHz |
| $g_{b1}$ | $2\pi \times -18.3$ MHz | $2\pi \times 19.3$ MHz |
| $g_{b2}$ | $2\pi \times -188.9$ MHz | $2\pi \times -195.5$ MHz |
| $\chi^{(1)}$ | $2\pi \times 0.9799$ MHz | $2\pi \times 1.0043$ MHz |
| $\chi^{(2)}$ | $2\pi \times 0.0109$ MHz | $2\pi \times 0.0121$ MHz |
| $\chi^{(ab)}$ | $2\pi \times 0.92$ MHz | $2\pi \times -0.85$ MHz |

with the the overlap functions $\beta_k \approx (E_{J_k}/8E_{C,k})^{1/4}/\sqrt{2}$.

We now tune the frequencies of the qubits close to resonance, $\Omega_{anc} \approx \omega_a$ and $\Omega_b \approx \omega_b$ and we transform to a rotating frame interaction picture with respect to $H_0 = \frac{\Omega_b}{2}\sigma_{z,b} + \frac{\Omega_{anc}}{2}\sigma_{z,anc} + \omega_a a^\dagger a + \omega_b b^\dagger b$. Here, second order time-dependent perturbation theory yields the interaction Hamiltonian $H_I = \frac{1}{\Delta}[\Xi^\dagger, \Xi]$ with $\Xi = a(g_{ab}b^\dagger + g_{a1}\sigma_{b,+}) + \sigma_{anc,-}(g_{b2}b^\dagger + g_{12}\sigma_{b,+})$ and $\Delta = \Omega_a - \Omega_b$ [5]. This gives us the Hamiltonian

$$H = \left(g_{b1} - \frac{g_{a1}g_{ab}}{\Delta} + \frac{g_{12}g_{b2}}{\Delta}\sigma_{anc,z}\right)(b^\dagger \sigma_{b,-} + \sigma_{b,+}b)$$
$$+ \left(g_{a2} + \frac{g_{b2}g_{ab}}{\Delta} - \frac{g_{12}g_{a1}}{\Delta}\sigma_{b,z}\right)(a^\dagger \sigma_{anc,-} + \sigma_{anc,+}a)$$
$$- \frac{g_{a1}^2}{\Delta}a^\dagger a \sigma_{b,z} + \frac{g_{b2}^2}{\Delta}b^\dagger b \sigma_{anc,z}. \tag{17}$$

At this point we want to saturate Qubit-$anc$ at a value conditioned upon the occupation in resonator $a$. To achieve this we implement a slight detuning $\delta = \omega_a - \Omega_{anc}$ between resonator $a$ and the ancillary qubit. Furthermore we require that $E_{J,anc}/E_{C,anc}$ is not too big, since this will allow a fast de-phasing of Qubit-$anc$ justifying its adiabatic elimination, such that we only have one qubit degree of freedom as in Eq. (1) and in $H_a$ of the main article.

We choose $\delta$ such that it is much larger than the effective coupling between the resonator and the qubit given by $g_{eff} = g_{a2} + \frac{g_{b2}g_{ab}}{\Delta} + \frac{g_{12}g_{a1}}{\Delta}$, and we assume a classical field amplitude $\alpha$ for resonator $a$. This gives us an effective damped Rabi-model where the steady state ancillary qubit inversion reads,

$$\left.(\rho_{ee} - \rho_{gg})\right|_{ss} = -1 + \zeta^{(1)}|\alpha|^2 - \zeta^{(2)}|\alpha|^4, \tag{18}$$

where $\zeta^{(1)} = g_{eff}^2/(\gamma^2 + 2\Delta^2)$ and $\zeta^{(2)} = 2g_{eff}^4/(\gamma^2 + 2\Delta^2)^2$ can be evaluated. This solution is reached much faster than the timescale of the field evolution in the resonators, and therefore we can safely utilize this steady state solution as the state of the qubit.

The task is now to choose parameters such that the switching condition is fulfilled [3],

$$g_{b1} \approx \frac{g_{a1}g_{ab}}{\Delta} + \frac{1}{2}\frac{g_{12}g_{b2}}{\Delta}, \tag{19}$$

($\sigma_{z,anc} \simeq -1/2$ in Eq.(17)), and at the same time to maximize the effective coupling,

$$\chi = \frac{g_{12}g_{b2}}{\Delta}(\zeta^{(1)} - \zeta^{(2)}a^\dagger a), \tag{20}$$

where we have returned to describe the resonator occupation by its operator expression. This is indeed possible, and we find $\chi^{(1)} = \frac{g_{12}g_{b2}}{\Delta}\zeta^{(1)}$ and $\chi^{(2)} = \frac{g_{12}g_{b2}}{\Delta}\zeta^{(2)}$ large enough that (1) gives rise to a strong coupling between resonator $b$ and qubit-1 when resonator $a$ is excited. The optimal quantities and the resulting coupling strengths are found numerically and displayed in Table 1. Controlling the capacitances precisely as specified in Tabel 1 may be difficult; however, once fabricated, one can tune $\Delta$, until Eq. (19) is fulfilled [6, 7]. Note as well that all effective couplings $g_{eff}$ and $\chi a^\dagger a$ are much smaller the anharmonicities, while the actual $g$'s are much smaller than $\Delta$, justifying our approximations.

As a last remark, we notice that Eq. (17) includes two dispersive terms. We ignore the term between Qubit-$b$ and resonator $a$, since they are rarely simultaneously excited (confirmed by our numerical simulations and shown in the section below on steady state solutions) and since $g_{a1}$ is small. The last term in (17) does, however, give rise to a cross-Kerr term,

$$H_{cross} = \chi^{(ab)}a^\dagger a b^\dagger b, \tag{21}$$

with $\chi^{(ab)} = \frac{g_{b2}^2}{\Delta}\zeta^{(1)}$. With the opposite choices for the detuning in the implementation of $H_a$ and $H_b$ (see Table 1), this term is effectively reduced.

## II. STEADY STATE SOLUTIONS

We are interested in the steady state solution for the Hamiltonian,

$$H = \chi\, a^\dagger a\, (b^\dagger \sigma_- + b\sigma_+), \tag{22}$$

where $a$ ($a^\dagger$) and $b$ ($b^\dagger$) is the annihilation (creation) operator for cavity-$a$ and cavity-$b$, while $\sigma_-$ and $\sigma_+$ are the ladder operators for a qubit. By incorporating only $H_a$ of the main article, we can estimate the lifetime of one of the memory states and infer that $H_b$ similarly protects the other memory state of the device (see discussion in the end of this section).

Both resonators of Eq. (22) are damped and driven by a classical fields and the qubit states have finite linetimes, so we describe the system by the master equation [8]

$$\begin{aligned}\frac{\partial \rho}{\partial t} &= i[\rho, H] + i\alpha[\rho, a + a^\dagger] + i\beta[\rho, b + b^\dagger] \\ &+ \kappa_a/2\,(2\,a\rho a^\dagger - a^\dagger a\rho - \rho a^\dagger a) \\ &+ \kappa_b/2\,(2\,b\rho b^\dagger - b^\dagger b\rho - \rho b^\dagger b) \\ &+ \gamma/2\,(2\,\sigma_-\rho\sigma_+ - \sigma_+\sigma_-\rho - \rho\sigma_+\sigma_-),\end{aligned} \tag{23}$$

with $\kappa_a$, $\kappa_b$ and $\gamma$ being the decay rate of the resonators and the qubit, while $\alpha$ and $\beta$ are the amplitudes of classical driving fields. In the following we will assume $\kappa_a = \kappa_b = \kappa$.

The system is expected to show bistable behaviour, revealed in the steady state solution by the possible correlation between the qubit and resonator field degrees of freedom. This motivates separating the master equation into four coupled equations – one for each term in the qubit density matrix [9],

$$\frac{\partial \rho_{\uparrow\uparrow}}{\partial t} = i\chi(\rho_{\uparrow\downarrow}a^\dagger a b^\dagger - a^\dagger a b \rho_{\downarrow\uparrow}) + i\alpha[\rho_{\uparrow\uparrow}, a + a^\dagger] + i\beta[\rho_{\uparrow\uparrow}, b + b^\dagger] + \mathcal{L}_a\rho_{\uparrow\uparrow} + \mathcal{L}_b\rho_{\uparrow\uparrow} - \gamma\rho_{\uparrow\uparrow} \tag{24}$$

$$\frac{\partial \rho_{\downarrow\downarrow}}{\partial t} = i\chi(\rho_{\downarrow\uparrow}a^\dagger a b - a^\dagger a b^\dagger \rho_{\uparrow\downarrow}) + i\alpha[\rho_{\downarrow\downarrow}, a + a^\dagger] + i\beta[\rho_{\downarrow\downarrow}, b + b^\dagger] + \mathcal{L}_a\rho_{\downarrow\downarrow} + \mathcal{L}_b\rho_{\downarrow\downarrow} + \gamma\rho_{\uparrow\uparrow} \tag{25}$$

$$\frac{\partial \rho_{\uparrow\downarrow}}{\partial t} = i\chi(\rho_{\uparrow\uparrow}a^\dagger a b - a^\dagger a b \rho_{\downarrow\downarrow}) + i\alpha[\rho_{\uparrow\downarrow}, a + a^\dagger] + i\beta[\rho_{\uparrow\downarrow}, b + b^\dagger] + \mathcal{L}_a\rho_{\uparrow\downarrow} + \mathcal{L}_b\rho_{\uparrow\downarrow} - \gamma/2\rho_{\uparrow\downarrow} \tag{26}$$

$$\frac{\partial \rho_{\downarrow\uparrow}}{\partial t} = i\chi(\rho_{\downarrow\downarrow}a^\dagger a b^\dagger - a^\dagger a b^\dagger \rho_{\uparrow\uparrow}) + i\alpha[\rho_{\downarrow\uparrow}, a + a^\dagger] + i\beta[\rho_{\downarrow\uparrow}, b + b^\dagger] + \mathcal{L}_a\rho_{\downarrow\uparrow} + \mathcal{L}_b\rho_{\downarrow\uparrow} - \gamma/2\rho_{\downarrow\uparrow}, \tag{27}$$

with $\mathcal{L}_k\rho = \kappa_k/2\,(2\,k\rho k^\dagger - k^\dagger k\rho - \rho k^\dagger k)$, $k = a, b$. The qubit coherence time is longer than the photon lifetime in the resonator but smaller than the timescale of storage in our device. This permits the approximate solution for the off-diagonal density-matrices

$$\rho_{\uparrow\downarrow} = \frac{2i\chi}{\gamma}(\rho_{\uparrow\uparrow}a^\dagger a b - a^\dagger a b \rho_{\downarrow\downarrow}) \tag{28}$$

$$\rho_{\downarrow\uparrow} = \frac{2i\chi}{\gamma}(\rho_{\downarrow\downarrow}a^\dagger a b^\dagger - a^\dagger a b^\dagger \rho_{\uparrow\uparrow}). \tag{29}$$

By substituting Eqs. (28) and (29) into Eqs. (24) and (25) we obtain

$$\frac{\partial \rho_{\uparrow\uparrow}}{\partial t} = \frac{-2\chi^2}{\gamma}(\rho_{\uparrow\uparrow}(a^\dagger a)^2 b b^\dagger - a^\dagger a b \rho_{\downarrow\downarrow} a^\dagger a b^\dagger - a^\dagger a b \rho_{\downarrow\downarrow} a^\dagger a b^\dagger + (a^\dagger a)^2 b b^\dagger \rho_{\uparrow\uparrow})$$
$$+ i\alpha[\rho_{\uparrow\uparrow}, a + a^\dagger] + i\beta[\rho_{\uparrow\uparrow}, b + b^\dagger] + \mathcal{L}_a\rho_{\uparrow\uparrow} + \mathcal{L}_b\rho_{\uparrow\uparrow} - \gamma\rho_{\uparrow\uparrow} \tag{30}$$

$$\frac{\partial \rho_{\downarrow\downarrow}}{\partial t} = \frac{-2\chi^2}{\gamma}(\rho_{\downarrow\downarrow}(a^\dagger a)^2 b^\dagger b - a^\dagger a b^\dagger \rho_{\uparrow\uparrow} a^\dagger a b - a^\dagger a b^\dagger \rho_{\uparrow\uparrow} a^\dagger a b + (a^\dagger a)^2 b^\dagger b \rho_{\downarrow\downarrow})$$
$$+ i\alpha[\rho_{\downarrow\downarrow}, a + a^\dagger] + i\beta[\rho_{\downarrow\downarrow}, b + b^\dagger] + \mathcal{L}_a\rho_{\downarrow\downarrow} + \mathcal{L}_b\rho_{\downarrow\downarrow} + \gamma\rho_{\uparrow\uparrow}. \tag{31}$$

To find the steady state of this set of equations we set $\frac{\partial \rho_{\downarrow\downarrow}}{\partial t} = 0 = \frac{\partial \rho_{\uparrow\uparrow}}{\partial t}$ and we assume the factorization [8]

$$\text{Tr}(a^s b^t \rho_{\uparrow\uparrow}(a^\dagger)^p(b^\dagger)^r) = (\alpha_\uparrow^*)^p(\beta_\uparrow^*)^r \alpha_\uparrow^s \beta_\uparrow^t P_\uparrow \tag{32}$$

$$\text{Tr}(a^s b^t \rho_{\downarrow\downarrow}(a^\dagger)^p(b^\dagger)^r) = (\alpha_\downarrow^*)^p(\beta_\downarrow^*)^r \alpha_\downarrow^s \beta_\downarrow^t P_\downarrow \tag{33}$$

with $\alpha_\uparrow, \alpha_\downarrow, \beta_\uparrow$ and $\beta_\downarrow$ complex numbers and $P_\uparrow$ and $P_\downarrow$ real probabilities. This factorization permits the field amplitudes in the resonators to acquire different values correlated with each other $(\alpha_\uparrow, \beta_\uparrow), (\alpha_\downarrow, \beta_\downarrow)$ and with the qubit state, and thus to reflect, even in steady state, the bistable character of the device. By normal ordering of the operators in Eq. (30) and (31) and by use of the above substitutions we obtain the equations

$$0 = -\frac{4\chi^2}{\gamma}\left((|\alpha_\uparrow|^4 + 2|\alpha_\uparrow|^2 + 1/2)(|\beta_\uparrow|^2 + 1)\alpha_\uparrow P_\uparrow - (|\alpha_\downarrow|^4 + 2|\alpha_\downarrow|^2)|\beta_\downarrow|^2 \alpha_\downarrow P_\downarrow\right) - i\alpha P_\uparrow - \frac{\kappa}{2}\alpha_\uparrow P_\uparrow - \gamma\alpha_\uparrow P_\uparrow \tag{34}$$

$$0 = -\frac{4\chi^2}{\gamma}\left((|\alpha_\downarrow|^4 + 2|\alpha_\downarrow|^2 + 1/2)|\beta_\downarrow|^2 \alpha_\downarrow P_\downarrow - (|\alpha_\uparrow|^4 + 2|\alpha_\uparrow|^2)(|\beta_\uparrow|^2 + 1)\alpha_\uparrow P_\uparrow\right) - i\alpha P_\downarrow - \frac{\kappa}{2}\alpha_\downarrow P_\downarrow + \gamma\alpha_\uparrow P_\uparrow \tag{35}$$

$$0 = -\frac{4\chi^2}{\gamma}\left((|\alpha_\uparrow|^4 + |\alpha_\uparrow|^2)(|\beta_\uparrow|^2 + 3/2)\beta_\uparrow P_\uparrow - (|\alpha_\downarrow|^4 + |\alpha_\downarrow|^2)|\beta_\downarrow|^2 \beta_\downarrow P_\downarrow\right) - i\beta P_\uparrow - \frac{\kappa}{2}\beta_\uparrow P_\uparrow - \gamma\beta_\uparrow P_\uparrow \tag{36}$$

$$0 = -\frac{4\chi^2}{\gamma}\left((|\alpha_\downarrow|^4 + |\alpha_\downarrow|^2)(|\beta_\downarrow|^2 + 1/2)\beta_\downarrow P_\downarrow - (|\alpha_\uparrow|^4 + |\alpha_\uparrow|^2)(|\beta_\uparrow|^2 + 2)\beta_\uparrow P_\uparrow\right) - i\beta P_\downarrow - \frac{\kappa}{2}\beta_\downarrow P_\downarrow + \gamma\beta_\uparrow P_\uparrow. \tag{37}$$

In these equations $P_\uparrow$ and $P_\downarrow$ denote the probability of being in the exited or the ground state of the qubit. These probabilities can be estimated from Eq. (30), and assuming that qubit relaxation is the primary decay mechanism, we get

$$\frac{P_\uparrow}{P_\downarrow} = \frac{\frac{4\chi^2}{\gamma^2}(|\alpha_\downarrow|^4 + |\alpha_\downarrow|^2)|\beta_\downarrow|^2}{1 + \frac{4\chi^2}{\gamma^2}\left((|\alpha_\uparrow|^4 + |\alpha_\uparrow|^2)(|\beta_\uparrow|^2 + 1)\right)} \tag{38}$$

which yields the qubit excitation probability

$$P_\uparrow = \frac{\frac{4\chi^2}{\gamma^2}|\alpha_\downarrow|^4|\beta_\downarrow|^2}{1 + \frac{4\chi^2}{\gamma^2}\left((|\alpha_\uparrow|^4 + |\alpha_\uparrow|^2)(|\beta_\uparrow|^2 + 1) + (|\alpha_\downarrow|^4 + |\alpha_\downarrow|^2)|\beta_\downarrow|^2\right)} \tag{39}$$

and $P_\downarrow = 1 - P_\uparrow$.

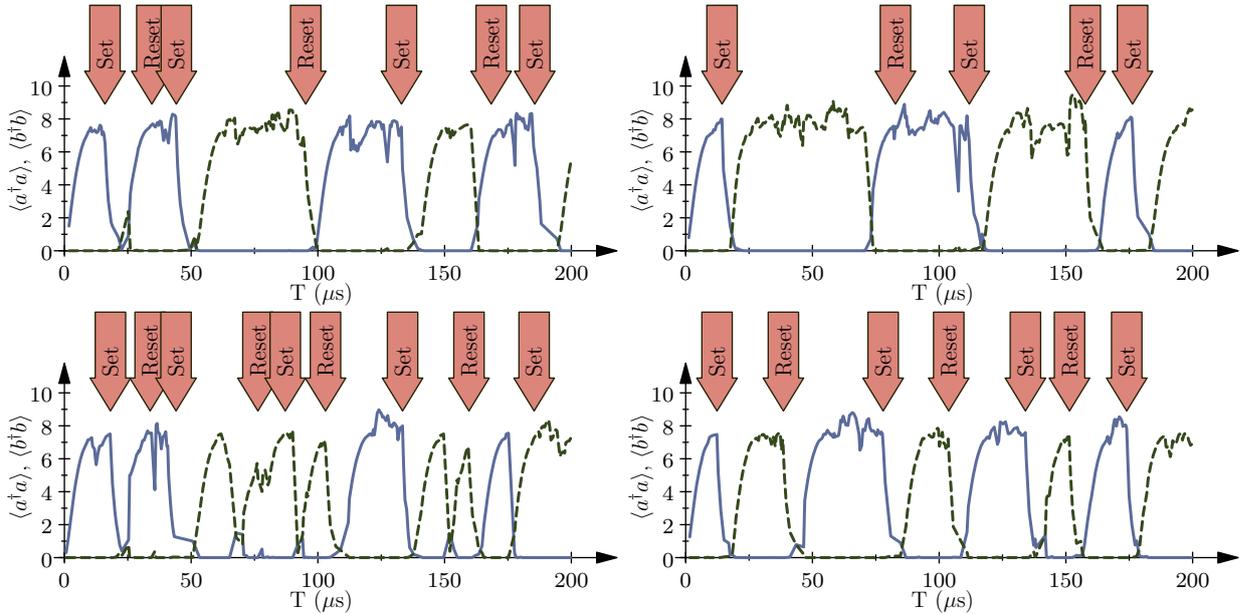

FIG. 2. (Color online) Single trajectories with Set and Reset pulses applied at the times shown . The solid (blue) curve shows $\langle a^\dagger a\rangle$ and the dashed (green) curve shows $\langle b^\dagger b\rangle$. The parameters used in the simulations are $(\chi_a^{(1)}, \chi_a^{(2)}, \chi_b^{(1)}, \chi_b^{(2)}, \chi^{(ab)}) = 2\pi \times (0.98, 0.011, 1.04, 0.012, 0.07)$ MHz and the resonator frequencies are $\omega_a = 2\pi \times 7$ GHz and $\omega_b = 2\pi \times 5$ GHz. We use a decay-rate of the cavities at $\kappa = 2\pi \times 0.1$ MHz and a lifetime of qubit-a and -b of $12\,\mu$s. For the transitors we assume $g_{ta/tb} = 2\pi \times 30$ MHz and lifetimes of $20\,\mu$s for both excited transistor levels.

Equations (34)-(37) can not be solved analytically, but we expect a pair of solutions with $\alpha(\beta)_\downarrow \sim -2i\alpha(\beta)/\kappa$ and $\beta(\alpha)_\downarrow \sim 0$. Numerically we confirm these two solutions, and we identify a third solution with all amplitudes very close to zero. However this solution is not observed in the simulations of the full dynamics of the Hamiltonian, and we classify it as an unstable solution of the effective mean field theory. We further notice that for "Solution 2" with $\beta_\downarrow$ close to $-2i\beta/\kappa$ we find $P_\uparrow$ to be much higher ($P_\uparrow \sim 10^{-1}$) than for "Solution 1" with $\alpha_\downarrow$ close to $-2i\alpha/\kappa$ ($P_\uparrow \sim 10^{-3}$) for typical parameters. We see that Solution 1 corresponds to the memory state with an excited resonator $a$ and a Rabi-splitting preventing excitation in resonator $b$, while Solution 2 blocks resonator $a$ by a small dispersive shift. Two similar solutions with $a \leftrightarrow b$ exist for $H_b$, given by Eq. (3) in the article, and by implementing both $H_a$ and $H_b$ we obtain two long lived memory states. We observe that one qubit is weakly excited in both memory states, but as confirmed by our simulations, it may decay without disrupting the feeding and the blocking of the resonators. Hence the memory time analysis based on either $H_a$ or $H_b$ describes also the coupled system.

### III. ADDITIONAL SIMULATIONS

We now turn our attention to the full dynamics of the device in the main article with both $H_a$ and $H_b$ active and with Set- and Reset-pulses applied. In Fig. 2 we supplement the trajectories in Fig. 3 of the main text to illustrate further features of the device. All parameters used here are the same as in the Letter.

In the upper left trajectory we see that after the first Set-pulse empties resonator $a$, when the transistor has decayed, resonator $b$ is occupied by less than two photons, which implies that in this trajectory we have a high probability for an unintended switch back to resonator $a$, which indeed happens. The system however recovers only after the next Reset-pulse. The upper right figure shows an unintended switch of the device around $t = 75\,\mu$s, and in the lower left figure we show a trajectory where both kinds of error occur. The lower right figure shows the ideal performance of the device, this time with a high frequency of switching pulses indicating a wider range of robustness. Accumulating the statistic of 30 simulations we find that the probability to have no error during $200\,\mu$s of evolution with 4–10 Set- and Reset-pulses is around 54 %, accounting both for the finite memory time ($\sim 347\,\mu$s) and errors occurring during

switches. Longer memory times and better transistor qubits may increase this probability significantly.

---



\end{document}